\documentclass[amssymb,amsmath,superscriptaddress,showpacs,nofootinbib]{revtex4}
\usepackage{amssymb}
\usepackage{bm}
\usepackage{aas_macros,braket}
\usepackage[dvipdfmx,final]{graphicx}
\input{colordvi.tex}
\begin{document}
\title{Analytic formulae of the CMB bispectra generated from
non-Gaussianity in the tensor and vector perturbations}
\author{Maresuke Shiraishi}
\affiliation{Department of Physics and Astrophysics, Nagoya University,
Aichi 464-8602, Japan}
\email{mare@a.phys.nagoya-u.ac.jp}
\author{Shuichiro Yokoyama}
\affiliation{Department of Physics and Astrophysics, Nagoya University,
Aichi 464-8602, Japan}
\author{Daisuke Nitta}
\affiliation{Department of Physics and Astrophysics, Nagoya University,
Aichi 464-8602, Japan}
\author{Kiyotomo Ichiki}
\affiliation{Department of Physics and Astrophysics, Nagoya University,
Aichi 464-8602, Japan}
\author{Keitaro Takahashi}
\affiliation{Department of Physics and Astrophysics, Nagoya University,
Aichi 464-8602, Japan}
\date{\today}

\begin{abstract}
 We present a complete set of formulae for calculating the
 bispectra of CMB temperature and polarization anisotropies generated
 from non-Gaussianity in the vector and tensor mode perturbations. In
 the all-sky analysis it is found that the bispectrum formulae for the
 tensor and vector-mode non-Gaussianity formally take complicated forms
 compared to the scalar mode one because the photon transfer functions
 in the tensor and vector modes depend on the azimuthal angle between
 the direction of the wave number vector of the photon's perturbation and that of the line of
 sight.  We demonstrate that flat-sky approximations remove this
 difficulty because this kind of azimuthal angle dependence apparently
 vanishes in the flat-sky limit. Through the flat-sky analysis, we also
 find that the vector or tensor bispectrum of $B$-mode polarization
 vanishes in the squeezed limit, unless the cosmological parity is
 violated at the nonlinear level.
\end{abstract}


\maketitle
\def\up{\;\raise1.0pt\hbox{$'$}\hskip-6pt\partial\;}
\def\down{\;\overline{\raise1.0pt\hbox{$'$}\hskip-6pt\partial\;}}
%~~~~~~~~~~~~~~~~~~~~~~~~~~~~~~~~~~~~~~~~~~~~~~~~~~~~~~~~~~~~~~~~~~~~~~~~~~~~
\section{Introduction}
Recently, the primordial non-Gaussianity of curvature perturbations
has been a focus of constant attention all over the world.  One of the
main reasons for attracting so much attention is that meaningful
measurement of this quantity will become observationally available in
the near future, which brings us valuable information about the dynamics
of inflaton.  Bispectrum (three point correlation functions) of the CMB
temperature anisotropies has been most commonly used to investigate primordial non-Gaussianity~\cite{Komatsu:2001rj,Bartolo/etal:2004}.  

As is well known, if the primordial curvature perturbation deviates from
the pure Gaussian statistics, then it produces the nonzero bispectrum
of the CMB temperature anisotropies.  However, there are a lot of
sources of the bispectrum that not only include the primordial
non-Gaussianity of the curvature perturbations but also the
nonlinearities of the Sachs-Wolfe
effect~\cite{Mollerach:1997up,Bartolo:2003bz,Bartolo:2005kv,Boubekeur:2009uk}
and the radiative
transfer~\cite{Bartolo:2006cu,Bartolo:2006fj,Pitrou:2008hy,Nitta:2009jp,Pitrou:2010sn},
cosmological
recombination~\cite{Khatri:2008kb,Bartolo:2008sg,Senatore:2008vi,Senatore:2008wk},
the nonlinear gravitational clustering of dark
matter~\cite{Pitrou:2008ak}, the cosmic
strings~\cite{Takahashi:2008ui,Hindmarsh:2009qk}, the magnetic
fields~\cite{Brown:2005kr,Seshadri:2009sy,Caprini:2009vk} and so on.
Hence, in order to evaluate the magnitude of the primordial
non-Gaussianity of the curvature perturbations precisely, it is very
important to identify these nonlinear effects.  

{In fact, these effects will also induce the tensor and vector-mode
perturbations, and the modes may generate more characteristic features
in the CMB angular spectra than in the scalar one. Some models have been
proposed in which vector modes are produced at the inflationary phase as
well as the scalar and tensor modes, by breaking the conformal
invariance at that phase \cite{1988PhRvD..37.2743T}. In such cases,
the non-Gaussian vector mode with an interesting amplitude could also be generated as in the scalar case. For example, if primordial
magnetic fields are considered, the magnetic stresses depend
quadratically on the primordial Gaussian magnetic field (PMF); hence,
their vector or tensor components of their bispectra also have finite
values. As the vector mode of the CMB transfer function sourced from
magnetic fields is dominant at small scale, one can expect that the vector-mode
bispectrum dominates there in the same manner as the power spectra
discussed in Refs.~\cite{Paoletti:2008ck, Kojima:2008rf,Shaw:2009nf}.
Therefore, if one adds the effects of the vector mode in constraining the
amplitude of PMF by using the CMB bispectra, one will obtain a tighter
bound than the current one as $\mathcal{O}(10) \rm nG$
\cite{Seshadri:2009sy,Caprini:2009vk}.  Furthermore, the cosmic strings
or the magnetic fields give more characteristic effects also in the
polarization spectra than in the temperature one \cite{Lewis:2004,
Lewis:2004sec, Seljak/etal:1997, Pogosian:2006hg}. Hence, for the
identification of the sources of the bispectrum, information of
temperature and polarization fluctuation generated from tensor and
vector-mode ones should be used, not only from the scalar mode
perturbations. However, there are not enough studies about their effects
yet.}

In this paper, we newly present the bispectrum formulae of the CMB
temperature and polarization anisotropies sourced from non-Gaussianity
in the tensor and vector-mode perturbations.   
First, we formulate all-sky bispectra generated from scalar, vector, and
 tensor modes and find that the bispectrum formulae for vector and
 tensor modes in all-sky analysis formally take complicated forms
 compared to the scalar mode case due to the dependence of the photon
 transfer functions on the azimuthal angle between the wave vector
 of photon perturbation $\bf k$ to the unit vector specifying the line of sight
 direction $\hat{n}$. Next, by using the flat-sky approximation, we simplify
 the equations of bispectra of the CMB anisotropies to solve the above
 difficulty because no azimuthal dependence arises in this limit. In
 addition, in our flat-sky formulae, we find that if the bispectra of
 $B$-mode polarization is a nonzero value, it infers the parity violation
 in the nonlinear sector.

{This paper is organized as follows: In Sec. \ref{sec:nonGauss}, 
we define the primordial non-Gaussianity from tensor
and vector perturbations. In Sec \ref{sec:all_bi}, we discuss the
formulae of the CMB bispectrum generated from tensor and vector
perturbations in the flat-sky analysis. In Sec \ref{sec:flat_bi}, we
explain the formulae of their bispectra in flat-sky
approximation. Finally, in Sec \ref{sec:sum}, we provide the
summary of this paper. In the Appendices, we derive the formulae of CMB
1-point function used in the discussion of Secs. \ref{sec:all_bi} and
\ref{sec:flat_bi}}

Throughout this paper, we assume that the Universe is spatially flat and use the definition of Fourier transformation:
\begin{eqnarray}
f({\bf x}) &\equiv& \int \frac{d^3 {\bf k}}{(2 \pi)^3} \tilde{f}({\bf k}) e^{i {\bf k} \cdot {\bf
 x}}~,\\
f({\bf \Theta}) &\equiv& \int \frac{d^2 {\boldsymbol \ell}}{(2 \pi)^2} \tilde{f}({\boldsymbol \ell}) e^{i {\boldsymbol \ell} \cdot {\bf
 \Theta}},
\end{eqnarray}
where $\bf \Theta$ and $\bf x$ are, respectively, 2D and 3D vectors in the configuration
space and $\boldsymbol \ell$ and $\bf k$ are, respectively, their Fourier conjugate variables. 

%%%%%%%%%%%%%%%%%%%%%%%%%%%%%%%%%%%%%%%
\section{Primordial power spectra and bispectra of the scalar, tensor and vector perturbations}
\label{sec:nonGauss}
%%%%%%%%%%%%%%%%%%%%%%%%%%%%%%%%%%%%%%%

In this section, we parametrize the primordial non-Gaussianity in the
tensor and vector perturbations.
As mentioned in the introduction,
in order to discuss the primordial non-Gaussianity,
the bispectrum of the fluctuations is commonly used. In this paper, we consider a general expression of the bispectrum of the tensor/vector perturbations which is given by
\begin{eqnarray}
\Braket{\xi^{s_1}({\bf k_1}) \xi^{s_2}({\bf k_2}) \xi^{s_3}({\bf
 k_3})} = (2\pi)^3 F^{s_1 s_2 s_3}(k_1, k_2, k_3) \delta^{(3)}({\bf
 k_1} + {\bf k_2} + {\bf k_3})~, \label{eq:general_PNG}
\end{eqnarray}
where $s_i$ expresses two helicity states: $\pm 1$ for a vector mode, $\pm 2$ for a tensor mode.
Here, for simplifying numerical calculation, we neglect the angular dependence of three wave number vectors in the bispectrum in the bispectrum $F$.
This expression includes the so-called ``squeezed'' or ``equilateral'' type of the non-Gaussianity~\cite{Babich:2004gb, Fergusson:2008ra}.

For example, the squeezed-type of the bispectrum is given as follow: As in Refs.~\cite{Bartolo/etal:2004,Verde/Wang/etal:2000, Komatsu:2001rj, Okamoto/Hu:2002},
the primordial power spectrum and bispectrum of the scalar curvature perturbations are introduced as
\begin{eqnarray}
\braket{\Phi_{\rm L}({\bf k_1}) \Phi_{\rm L}({\bf k_2})} &=& (2 \pi)^3 P_\Phi(k_1)
 \delta^{(3)}({\bf k_1} + {\bf k_2})~, \\
\braket{\Phi_{\rm L}({\bf k_1}) \Phi_{\rm L}({\bf k_2}) \Phi_{\rm NL}({\bf
 k_3})} &=& (2 \pi)^3 P_\Phi(k_1) P_\Phi(k_2) 2 f_{\rm NL} \delta^{(3)}({\bf k_1} +
 {\bf k_2} + {\bf k_3}), 
\end{eqnarray}
where $f_{\rm NL}$ is the nonlinear parameter of the scalar
perturbation and $\Phi({\bf k})$ denotes the Fourier component of the
primordial curvature perturbation which is decomposed into Gaussian and
non-Gaussian part as
\begin{eqnarray}
\Phi({\bf x}) &\equiv& \Phi_{\rm L}({\bf x}) + \Phi_{\rm NL}({\bf x})~, \\
\Phi_{\rm NL}({\bf x}) &\equiv& f_{\rm NL} [\Phi_{\rm L}({\bf x})^2 - \braket{\Phi_{\rm L}({\bf x})^2}]~,
\end{eqnarray}
in real space. 

This parametrization can be readily extended to the tensor and vector cases.  In
contrast to the scalar perturbation, Fourier modes of tensor and vector
perturbations have two independent polarizations.  For the
convenience of calculating the CMB power spectrum as discussed in
 Refs.~\cite{Zaldarriaga/Seljak:1997, Weinberg:2008, Brown:2008}, we use two helicity
states ($ \pm 1$ for a vector mode, $\pm 2$ for a tensor mode) to
decompose the initial stochastic fields, $\xi^{s}$, where $s$ represents
a helicity state.  We apply this description to the definition of the
initial non-Gaussianity of the tensor and vector perturbations
as
\footnote{ One can easily include the scalar mode into our notation by
considering $s=0$ initial stochastic field.  For such a case, $\xi^0=\Phi$
and ${1 \over 2}f_{S,00}^{0} = f_{\rm NL}$ where the index $S$ is used for the
scalar mode.  }
\begin{eqnarray}
\xi^s({\bf x}) &\equiv& \xi^s_{{\rm L}}({\bf x}) + \xi^s_{{\rm NL}}({\bf x})~, \label{eq:def_xi}\\
\xi^{s_k}_{{\rm NL}}({\bf x}) &\equiv& \frac{1}{2}f^{s_k}_{Z,{s_i s_j}} 
\left[ \xi^{s_i}_{{\rm L}}({\bf x}) \xi^{s_j}_{{\rm L}}({\bf x}) -
 \braket{ \xi^{s_i}_{{\rm L}}({\bf x}) \xi^{s_j}_{{\rm L}}({\bf x}) }
\right]~. 
\label{eq:def_xi_NL}
\end{eqnarray}
where
the index $Z = T$ is used for the tensor mode 
($s_i, s_j, s_k = \pm 2$)
and $=V$ for the vector mode 
($s_i, s_j, s_k = \pm 1$).
Here, we have introduced new nonlinear parameters for the tensor and vector perturbations denoted by $f^{s_k}_{Z,{s_i} {s_j}}$.
These three indices, $s_i$, $s_j$, $s_k$, allow the correlation
between each field of the different helicity states in the nonlinear level. Because of the symmetry, we have 
$f_{Z,s_i s_j}^{s_k} = f_{Z,s_j s_i}^{s_k}$.
By using these expressions, the primordial power spectra and bispectra
of the tensor and vector perturbations are expressed as
\footnote{For the scalar mode, $P_{S}(k)/2 = P_{\Phi}(k)$.}
\begin{eqnarray}
\braket{\xi^{s_1}_{{\rm L}}({\bf k_1}) \xi^{s_2}_{{\rm L}}({\bf k_2})} 
&=& (2 \pi)^3
 \frac{P_{Z}(k_1)}{2} \delta_{{s_1}{s_2}} \delta^{(3)}({\bf k_1} + {\bf k_2})~,  \\
\braket{\xi^{s_1}_{{\rm L}}({\bf k_1}) \xi^{s_2}_{{\rm L}}({\bf k_2}) \xi^{s_3}_{{\rm NL}}({\bf k_3})} &=& (2 \pi)^3 \frac{P_{Z}(k_1)}{2} \frac{P_{Z}(k_2)}{2}
 f^{s_3}_{Z,{s_1} {s_2}} \delta^{(3)}({\bf k_1} + {\bf k_2} + {\bf k_3})~.
\end{eqnarray}
Then, the squeezed type of the non-Gaussianity  can be expressed as
\begin{eqnarray}
F^{s_1 s_2 s_3}(k_1, k_2, k_3) 
= \left( \frac{P_{Z}(k_1)}{2} \frac{P_{Z}(k_2)}{2}
 f^{s_3}_{Z,{s_1} {s_2}} + 2 \ {\rm perms.}\right)~. \label{eq:local_PNG}
\end{eqnarray} 

In the following discussion, 
we use the general expression (\ref{eq:general_PNG}) as the bispectra of the tensor and vector perturbations without specifying the type of the non-Gaussianity.

%#################################################
%############
\section{CMB bispectrum in the all-sky analysis} \label{sec:all_bi}

%%%%%%%%%%%%%%%%%%%%%%%%%%%%%%%%%%%%%%%%%%%%%%%%%%%%%%%%
In this section, we derive the formulae of the CMB bispectra sourced
from tensor and vector perturbations on the full sky.
The primordial perturbations in the scalar, vector, and tensor sectors
introduced in the previous section are
transferred through the primordial plasma to the CMB epoch and observed
in the CMB temperature and polarization fluctuations.
In the all-sky analysis, the CMB spin-0 temperature field $I$ and spin-2
polarization fields $Q,U$, are expanded by spin-weighted spherical
harmonics \cite{Zaldarriaga/Seljak:1997, Lewis:2004}.  Following the
usual manner, we convert the $Q \pm iU$ fields into spin-0 $E$ and $B$ fields by using the ``spin raising operator'' and `` spin lowering operator'' as Eqs. (\ref{eq:form_all_aelm}) and (\ref{eq:form_all_ablm}). Their radiative transfer
functions are shown in Appendix \ref{appen:all_al}.
\subsection{Scalar mode case}
%--- scalar perturbation bispectrum
First, we give a brief review of the CMB bispectrum sourced from scalar perturbation.
For the scalar case, the CMB bispectrum can be written as~\cite{Bartolo/etal:2004}  
\begin{eqnarray}
\braket{a^{(S)}_{X,\ell_1 m_1} a^{(S)}_{X,\ell_2 m_2} a^{(S)}_{X,\ell_3 m_3}} 
= \mathcal{G}^{m_1 m_2 m_3}_{\ell_1 \ell_2 \ell_3} b_{X, \ell_1 \ell_2 \ell_3}^{(S)}~, \label{eq:all_bi_S} 
\end{eqnarray}
where the index $(S)$ means that a source of the CMB fluctuation is the scalar perturbation and
the index $X$ denotes the temperature ($I$), the $E$-mode polarization ($E$) and the $B$-mode polarization ($B$)\footnote{
Of course, scalar perturbation contributes only to the $E$-mode polarization and not to the $B$-mode one.
In this paper, we use this index also for the tensor and vector modes which contribute not only to the $E$ mode but also the $B$ mode.
}.
Here,
$\mathcal{G}^{m_1 m_2 m_3}_{\ell_1 \ell_2 \ell_3}$ is the Gaunt integral given by
\begin{eqnarray}
 \mathcal{G}^{m_1 m_2 m_3}_{\ell_1 \ell_2 \ell_3} &\equiv& \int d\Omega_y Y_{\ell_1
  m_1}(\Omega_y) Y_{\ell_2 m_2}(\Omega_y) Y_{\ell_3 m_3}(\Omega_y)~, \label{eq:gaunt}
\end{eqnarray}
and $b_{X, \ell_1 \ell_2 \ell_3}^{(S)}$ is the scalar reduced bispectrum formulated as 
\begin{eqnarray}
b_{X, \ell_1 \ell_2 \ell_3}^{(S)}
 = \int_0^\infty y^2 dy 
\left[ \prod_{i=1}^3 \frac{2}{\pi} \int_0^\infty k^2_i d k_i j_{\ell_i}(k_i y) 
\mathcal{T}^{(S)}_{X, \ell_i}(k_i) \right] F^{000}(k_1, k_2, k_3) ~. \label{eq:all_red_bi_S}
\end{eqnarray}
Here $\mathcal{T}^{(S)}_{X,\ell_i}(k_i)$ is the time-integrated transfer function of the scalar perturbation as shown in Eqs. (\ref{eq:all_almst}) and (\ref{eq:all_almse}), and $j_\ell(x)$ is the spherical Bessel function.

%==================
\subsection{Tensor and vector-mode case}

%--- tensor perturbation bispectrum
Let us follow the above formulation for the tensor and vector cases.
Tensor and vector 1-point functions are explicitly given as
Eqs. (\ref{eq:all_alm_general}), (\ref{eq:all_almtt}) - (\ref{eq:all_almvb}) in Appendix B.
%D-matrix 
From those equations, one may wonder why tensor and vector 1-point
functions depend on the spin-weighted spherical harmonics
${}_{s}Y_{lm}(\Omega_k)$ although $I,E$ and $B$ modes are spin-0 fields.
This dependence arises as a consequence of calculating the transfer
function in the arbitrary direction of the wave vector $\bf k$.  As
discussed in detail in Appendix \ref{appen:all_al}, the transfer
function for the arbitrary $\bf k$ is written with the Wigner $D$ matrix
\cite{Weinberg:2008, Okamoto/Hu:2002, Goldberg/etal:1967} under the
rotational transformation of $\bf k$ from a particular direction
(e.g., $z$ direction)
to an arbitrary direction.  This $D$ matrix can be transcribed into
${}_{ s}Y_{lm}(\Omega_k)$ as Eq. (\ref{eq:d-matrix}).  As we will show
in the following discussion, because of this spin-weighted spherical
harmonics ${}_{s}Y_{lm}(\Omega_k)$, the CMB bispectra sourced from the
tensor and vector modes on each angular momentum, $\ell$, depends on the
sum of the reduced bispectrum over all angular momenta 
\footnote{ This
complexity does not occur for the CMB 2-point power spectra sourced
even from the tensor and vector modes and as is well known all CMB power
spectra can be described as
\begin{eqnarray}
\langle a^{(Z)*}_{X',\ell' m'}a^{(Z)}_{X,\ell m}\rangle = C_{X' X, \ell}^{(Z)} \delta_{\ell'\ell}\delta_{m'm}~.
\end{eqnarray}
}. 

For example, let us consider the CMB temperature fluctuation sourced from the tensor perturbation which
has the spin-$2$ spherical harmonics as
\begin{eqnarray}
a_{I,\ell m}^{(T)} \supset {}_{-2}Y_{\ell m}(\Omega_{k}) \xi^{+2}({\bf k}),~{}_{+2}Y_{\ell m}(\Omega_k)\xi^{-2}({\bf k})~.
\label{eq:alm_tens_temp}
\end{eqnarray}
Here we consider that the bispectrum of tensor-temperature fluctuations
can be sourced from the non-Gaussianity of the primordial tensor
perturbations which is characterized by the primordial bispectrum given
by Eq.~(\ref{eq:general_PNG}) and hence we can easily find
\begin{eqnarray}
\langle a_{I,\ell_1 m_1}^{(T)} a_{I,\ell_2 m_2}^{(T)} a_{I, \ell_3 m_3}^{(T)} \rangle 
\supset
\delta^{(3)}({\bf k}_1 + {\bf k}_2 + {\bf k}_3)~.
\end{eqnarray}
By making use of the expansion of 3D Dirac delta function given by
\begin{eqnarray}
\delta^{(3)}({\bf k}_1 + {\bf k}_2 + {\bf k}_3) 
&=& \int {d^3{\bf y} \over (2\pi)^3} e^{i({\bf k}_1 + {\bf k}_2 + {\bf k}_3)\cdot {\bf y}} \nonumber \\ 
&=& \int^{\infty}_0 y^2 dy \int d\Omega_y \prod_{i=1}^3 2 \sum_{\ell_i'
 m_i'} i^{\ell_i'} j_{\ell_i'} (k_i y) Y^*_{\ell_i' m_i'}(\Omega_y)
 Y_{\ell_i' m_i'}(\Omega_{k_i})~,
\label{Eq:3d_delta_expansion}
\end{eqnarray}
and Eq.~(\ref{eq:alm_tens_temp}),
a part of the bispectrum of tensor-temperature fluctuations can be expressed as
\begin{eqnarray}
\langle a_{I,\ell_1 m_1}^{(T)} a_{I,\ell_2 m_2}^{(T)} a_{I, \ell_3 m_3}^{(T)} \rangle 
\supset \sum_{\ell'_1 \ell'_2 \ell'_3 \atop m'_1 m'_2 m'_3}\mathcal{G}^{m'_1 m'_2 m'_3}_{\ell'_1 \ell'_2 \ell'_3}
\prod_{i=1}^{3} \int d\Omega_{k_i} Y_{\ell'_i m'_i}(\Omega_{k_i}){}_{-2}Y^*_{\ell_i m_i}(\Omega_{k_i})~.
\end{eqnarray}
Although the bispectrum of the scalar-temperature fluctuation (and
$E$-mode polarization induced from scalar-type perturbation) is derived
in the same manner, the orthogonality of the spin-$0$ spherical harmonic
functions gives us quite simple expression of Eq.~(\ref{eq:all_bi_S})
whose form is the Gaunt integral multiplied by the scalar reduced
bispectrum.  However, as seen in the above expression in the tensor case
(also the vector case) the CMB bispectra on each $\ell$ depends on the
sum of the reduced bispectrum over all angular momenta $\ell'$ as
Eq. (\ref{eq:all_bi_TV}) in contrast to the scalar case such as
Eq. (\ref{eq:all_bi_S}), due to the nonorthogonality of the $\theta_k$
dependence between $Y_{\ell' m'}$ and ${}_{s}Y_{\ell m}$ ($s=\pm 1$ or $\pm 2$).
One may think that this complexity would be evaded once the
plane wave could be expanded using spin-weighted spherical harmonics,
rather than Eq. (\ref{Eq:3d_delta_expansion}). In this paper, however,
instead of pursuing this possibility we will use the flat-sky
approximation to evade this difficulty as we shall show below.

Thus, the bispectrum formulae of the CMB fluctuations sourced from tensor and vector modes are, respectively, given by 
\begin{eqnarray}
\braket{a^{(Z)}_{X,\ell_1 m_1} a^{(Z)}_{X,\ell_2 m_2} a^{(Z)}_{X,\ell_3 m_3}}
&=& \sum_{\ell'_1,\ell'_2,\ell'_3} i^{\ell'_1+\ell'_2+\ell'_3}
\mathcal{G}^{m_1 m_2 m_3}_{\ell'_1 \ell'_2 \ell'_3} 
\int_0^\infty y^2 dy \left[ \prod_{i=1}^3 \frac{2}{\pi} (-i)^{\ell_i} \int_0^\infty k^2_i d k_i j_{\ell'_i}(k_i y) \mathcal{T}^{(Z)}_{X, \ell_i}(k_i) \right]
\nonumber \\
&& \times
\sum_{s_1 s_2 s_3} {\rm sgn}(s_1)^{s_1 + x_1} {\rm sgn}(s_2)^{s_2 + x_2} {\rm sgn}(s_3)^{s_3 + x_3} F^{s_1 s_2 s_3}(k_1, k_2, k_3) \mathcal{Y}^{(Z)}_{\ell \ell' m}(s) ~,
 \label{eq:all_bi_TV}
\end{eqnarray}
with 
\begin{eqnarray}
\mathcal{Y}^{(Z)}_{\ell \ell' m}(s)
\equiv \prod_{i=1}^{3}
\int d\Omega_{k_i}  Y_{\ell'_i m_i}(\Omega_{k_i}) {}_{- s_i}Y_{\ell_i m_i}^*(\Omega_{k_i})~.
\end{eqnarray}
Here $\mathcal{T}^{(Z)}_{X, \ell_i}(k_i)$ is the time-integrated transfer function generated from vector $(Z=V)$ or tensor $(Z=T)$ perturbation as described in Eqs. (\ref{eq:all_almtt}) - (\ref{eq:all_almvb}) and $x$ is the index: $x \equiv 0$ for $X = I,E$ and $x \equiv 1$ for $X = B$.

As we have mentioned before, 
in the tensor and vector cases, due to the nonorthogonality of
the $\theta_k$ dependence between ${}_{s}Y_{\ell m}$ and $Y_{\ell' m'}$,
the CMB bispectra on each $\ell$ depend on
the sum of the reduced bispectrum over all angular momenta $\ell'$
as Eq. (\ref{eq:all_bi_TV}) in contrast to the scalar case such as Eq. (\ref{eq:all_bi_S}).
For this complexity, the numerical calculations of the tensor and vector
bispectra take much longer time than that of the scalar one.  However,
this problem can be evaded by using the flat-sky approximation as shown
in the next section.

%%%%%%%%%%%%%%%%%%%%%%%%%%%%%%%%%%%%%%%%%%%%%
\section{CMB bispectra in the flat-sky analysis} \label{sec:flat_bi}

Here, we explain the formulation of the CMB bispectrum
sourced from tensor and vector perturbations by using the flat-sky
approximation as mentioned in Refs.~\cite{Zaldarriaga/Seljak:1997,
Seljak:1997, Boubekeur:2009uk, Pitrou:2008ak}.
% flat sky difinition
The flat-sky approximation uses the (2D) plane wave expansion of the CMB fluctuation instead of the spherical harmonics one, and it is valid if we restrict observed direction
 $\hat{n}$ only close to the
$z$ axis.  As confirmed in Ref.~\cite{Zaldarriaga/Seljak:1997}, the flat-sky
power spectra of $E$- and $B$-mode polarizations sourced from the
primordial tensor perturbations are in good agreement with the all-sky
ones for $\ell \gtrsim 40$.  In Ref.~\cite{Boubekeur:2009uk}, the validity
of the flat-sky analysis is also shown in the calculation of the
temperature bispectra generated from the Sachs-Wolfe term by evaluating
the convergence of the modified Bessel function.  In addition, in
 Ref.~\cite{Pitrou:2008ak}, the consistency between the flat-sky result and all-sky one in the calculation of the scalar-temperature power spectrum and bispectrum are discussed.

Based on these studies, we have also compared the all-sky power spectra
with the flat-sky ones for the $I,E,B$ modes from the tensor and vector
perturbations and found their consistencies at $\ell \gtrsim 40$.  We
have also compared all-sky and flat-sky temperature bispectra induced
from scalar-type perturbations, and confirmed that the flat-sky
approximation is also applicable in the calculation of the bispectrum
for the angular scales where the flat-sky power spectrum is a good
approximation of the all-sky power spectrum.  From these considerations,
even if we can not compare the all-sky bispectra with the flat-sky ones
in the tensor and vector modes due to the difficulties discussed in the
previous section, we can regard the flat-sky bispectra from the tensor and
vector perturbations as good approximations for $\ell \gtrsim 40$.

\subsection{Scalar bispectra in the flat-sky analysis}
%------- scalar reduced bispectrum
As described in Refs.~\cite{Komatsu:2001rj, Bartolo/etal:2004}, in the flat-sky approximation the scalar bispectrum Eq. (\ref{eq:all_bi_S}) is modified as
\begin{eqnarray}
\braket{a^{(S)}_{X}({\boldsymbol \ell}_1) a^{(S)}_{X}({\boldsymbol \ell}_2) a^{(S)}_{X}({\boldsymbol \ell}_3)}
= (2\pi)^2 \delta^{(2)}({\boldsymbol \ell}_1 + {\boldsymbol \ell}_2 +
{\boldsymbol \ell}_3) b_X^{(S)}(\ell_1,\ell_2,\ell_3)~, \label{eq:flat_bi_S}
\end{eqnarray}
Since $\mathcal{G}^{m_1 m_2 m_3}_{\ell_1 \ell_2 \ell_3} \approx (2\pi)^2
\delta^{(2)}({\boldsymbol \ell}_1 + {\boldsymbol \ell}_2 + {\boldsymbol
\ell}_3)$, Eq. (\ref{eq:flat_bi_S}) indicates 
$b^{(S)}_{X, \ell_1 \ell_2 \ell_3} \approx b_X^{(S)}(\ell_1,\ell_2,\ell_3)$.  A detailed derivation of 1-point
functions $a^{(Z)}_X ({\boldsymbol \ell})$ is 
presented in Appendix C. The scalar reduced bispectra are formulated,
by using Eqs. (\ref{eq:flat_alst}) and (\ref{eq:flat_alse}), as
\begin{eqnarray}
b_X^{(S)}(\ell_1,\ell_2,\ell_3)
= \int^{\infty}_{-\infty} y^2 dy 
\left[ \prod_{i=1}^3
\int_0^{\tau_0} d\tau_i \int_{\ell_i/D_i}^{\infty} \frac{dk_i}{2\pi} 
g_{X}^{(S)}(\ell_i, k_i, \tau_i, y)  \right] 
F^{000}(k_1, k_2, k_3)  ~, \label{eq:flat_red_bi_S}
\end{eqnarray}
where $D_i \equiv \tau_0 - \tau_i$ and the scalar $g$ functions are described as
\begin{eqnarray}
g_{I}^{(S)}(\ell, k, \tau, y) &=& 
 S_I^{(S)}(k,\tau) \frac{k}{\sqrt{k^2 - (\ell/D)^2}} \frac{2}{D^2} 
\cos \left[ \sqrt{k^2 - (\ell/D)^2} (y-D)\right]~, \\
%==============
g_{E}^{(S)}(\ell, k, \tau, y) &=&  
 S_P^{(S)}(k,\tau) \frac{k}{\sqrt{k^2 - (\ell/D)^2}} 
\left( \frac{\ell}{kD} \right)^2
\frac{2}{D^2} 
\cos \left[ \sqrt{k^2 - (\ell/D)^2} (y-D)\right]~.
\end{eqnarray}
Here $S_I^{(S)}(k,\tau)$ and $S_P^{(S)}(k,\tau)$ are the scalar-type source functions of the temperature and polarization fluctuations as mentioned in Appendix. \ref{appen:transfer}
%=====================
\subsection{Tensor and vector bispectra in the flat-sky analysis}
Let us consider the tensor-temperature bispectrum in the flat-sky analysis.
By using Eq. (\ref{eq:flat_altt}), a component of the flat-sky bispectrum of tensor-temperature mode is written as
\begin{eqnarray}
\braket{a^{(T)}_{I}({\boldsymbol \ell}_1) a^{(T)}_{I}({\boldsymbol \ell}_2) a^{(T)}_{I}({\boldsymbol \ell}_3)} 
&=&  \left[ \prod_{i=1}^3 \int_0^{\tau_0} d\tau_i \int_{-\infty}^{\infty} \frac{d{k_i}_z}{2\pi} S_I^{(T)}
(k_i = \sqrt{{k_i}_z^2 + (\ell_i / D_i)^2},\tau_i) \frac{\ell_i^2}{({k_i}_z D_i)^2 + \ell_i^2}\frac{1}{D_i^2}e^{-i {k_i}_z D_i} \right] \nonumber \\
&& \times \sum_{s_1, s_2, s_3 = \pm 2} 
F^{s_1 s_2 s_3} ( \sqrt{{k_1}_z^2 + (\ell_1 / D_1)^2}, \sqrt{{k_2}_z^2 + (\ell_2 / D_2)^2}, \sqrt{{k_3}_z^2 + (\ell_3 / D_3)^2} )  
 \nonumber \\
&& \times (2 \pi)^3  \delta^{(2)}(\frac{{\boldsymbol \ell}_1}{D_1} + \frac{{\boldsymbol \ell}_2}{D_2} + \frac{{\boldsymbol \ell}_3}{D_3}) \delta({k_1}_z + {k_2}_z + {k_3}_z)~. 
\end{eqnarray}
By using the expansion of the 1D Dirac
delta function and the approximation of 2D Dirac delta function as
  \begin{eqnarray}
\delta({k_1}_z + {k_2}_z + {k_3}_z) &=& \int_{-\infty}^{\infty}
 \frac{dy}{2\pi} e^{i({k_1}_z + {k_2}_z + {k_3}_z)y} \label{eq:delta1}~, \\
\delta^{(2)}(\frac{{\boldsymbol \ell}_1}{D_1} + \frac{{\boldsymbol \ell}_2}{D_2} + \frac{{\boldsymbol \ell}_3}{D_3}) &=& D_1^2 \delta^{(2)}({\boldsymbol \ell}_1 + {\boldsymbol \ell}_2 + {\boldsymbol \ell}_3 
+ \frac{D_1-D_2}{D_2}{\boldsymbol \ell}_2 + \frac{D_1-D_3}{D_3}{\boldsymbol \ell}_3) \nonumber \\
&\approx& D_1^2 \delta^{(2)}({\boldsymbol \ell}_1 + {\boldsymbol \ell}_2 + {\boldsymbol \ell}_3)~, \label{eq:delta2}
\end{eqnarray}
we can derive the simple form of the tensor-temperature bispectrum in the flat limit as 
\begin{eqnarray}
\braket{a_{I}^{(T)}({\boldsymbol \ell}_1) a_{I}^{(T)}({\boldsymbol \ell}_2) a_{I}^{(T)}({\boldsymbol \ell}_3)}
&\approx& (2\pi)^2 \delta^{(2)}({\boldsymbol \ell}_1 + {\boldsymbol \ell}_2 + {\boldsymbol \ell}_3) 
\int^{\infty}_{-\infty} y^2 dy 
\left[\prod_{i=1}^3 \int_0^{\tau_0} d\tau_i \int_{\ell_i/D_i}^{\infty} \frac{dk_i}{2\pi} g_{I}^{(T)}(\ell_i, k_i, \tau_i, y) \right] \nonumber \\ 
&& \times 
\sum_{s_1, s_2, s_3 = \pm 2} F^{s_1 s_2 s_3} (k_1, k_2, k_3)~  . \label{eq:flat_bi_TI_derive} 
\end{eqnarray} 
The approximation of Eq. (\ref{eq:delta2}) is valid because the
bispectra are suppressed when the triangle in the ${\boldsymbol
\ell}$-space does not close as discussed in Ref.~\cite{Boubekeur:2009uk}.
In Eq. (\ref{eq:flat_bi_TI_derive}), we use the approximation $(D_1/y)^2 \approx 1$, which is valid because the integrand has large value for $D_1 \sim y \sim \tau_0$.
Similar to the discussion in the previous section, as the other tensor
bispectra and the vector bispectra can be derived in the same manner,
these bispectra can be written by the same form as the scalar
bispectra~of Eq. (\ref{eq:flat_bi_S}) which can be written as a 2D Dirac delta
function multiplied by the reduced bispectra;
\begin{eqnarray}
\braket{a_{X}^{(Z)}({\boldsymbol \ell}_1) a_{X}^{(Z)}({\boldsymbol \ell}_2) a_{X}^{(Z)}({\boldsymbol \ell}_3)}
&=& (2\pi)^2 \delta^{(2)}({\boldsymbol \ell}_1 + {\boldsymbol \ell}_2 +
{\boldsymbol \ell}_3) b_X^{(Z)}(\ell_1,\ell_2,\ell_3)~, \label{eq:flat_bi_TV}
\end{eqnarray}
where the tensor or vector reduced bispectrum is expressed as
\begin{eqnarray}
b_X^{(Z)}(\ell_1,\ell_2,\ell_3)
&=& \int^{\infty}_{-\infty} y^2 dy 
\left[\prod_{i=1}^3 \int_0^{\tau_0} d\tau_i \int_{\ell_i/D_i}^{\infty} \frac{dk_i}{2\pi} g_{X}^{(Z)}(\ell_i, k_i, \tau_i, y) \right] \nonumber \\
&&\times \sum_{s_1, s_2, s_3} 
{\rm sgn}(s_1)^{x_1} {\rm sgn}(s_2)^{x_2} {\rm sgn}(s_3)^{x_3} 
F^{s_1 s_2 s_3} (k_1, k_2, k_3)
~. \label{eq:flat_red_bi_TV}
\end{eqnarray}
Tensor $g$ functions are written as
\begin{eqnarray}
%---- tensor
g_{I}^{(T)}(\ell, k, \tau, y) &=& 
 S_I^{(T)}(k,\tau) \frac{k}{\sqrt{k^2 - (\ell/D)^2}} \left(\frac{\ell}{kD}\right)^2
\frac{2}{D^2} 
\cos \left[ \sqrt{k^2 - (\ell/D)^2} (y-D)\right]~, \\
%=====
g_{E}^{(T)}(\ell, k, \tau, y) &=&  S_P^{(T)}(k,\tau)  \frac{k}{\sqrt{k^2 - (\ell/D)^2}}
\left[ 2 - \left(\frac{\ell}{kD}\right)^2 \right]
\frac{2}{D^2} \cos \left[ \sqrt{k^2 - (\ell/D)^2} (y-D)\right]~, \\
%========
g_{B}^{(T)}(\ell, k, \tau, y) &=& -  S_P^{(T)}(k,\tau)  
\frac{4}{D^2} \sin \left[ \sqrt{k^2 - (\ell/D)^2} (y-D)\right]~. \label{eq:parity_t}
\end{eqnarray}
Vector $g$ functions are also described as
\begin{eqnarray}
%%%%% vector 
g_{I}^{(V)}(\ell, k, \tau, y) &=& i~ S_I^{(V)}(k,\tau) \frac{\ell}{\sqrt{(kD)^2 - \ell^2}}
\frac{2}{D^2} 
\cos \left[ \sqrt{k^2 - (\ell/D)^2} (y-D)\right] ~,\\
%===================
g_{E}^{(V)}(\ell, k, \tau, y) &=& - i~ S_P^{(V)}(k,\tau) \left(\frac{\ell}{kD}\right) 
 \frac{2}{D^2} 
\sin \left[ \sqrt{k^2 - (\ell/D)^2} (y-D)\right] ~,\\
%============
g_{B}^{(V)}(\ell, k, \tau, y) &=& - i~ S_P^{(V)}(k,\tau) 
\frac{\ell}{\sqrt{(kD)^2 - \ell^2}}
 \frac{2}{D^2} 
\cos \left[ \sqrt{k^2 - (\ell/D)^2} (y-D)\right] ~. \label{eq:parity_v}
\end{eqnarray}
Here $S_I^{(Z)}(k,\tau)$ and $S_P^{(Z)}(k,\tau)$ are the $Z$-type source functions of the temperature and polarization fluctuations as mentioned in Appendix. \ref{appen:transfer}.

By comparing Eq. (\ref{eq:flat_bi_TV}) to Eq. (\ref{eq:flat_bi_S}), we find
that the tensor and vector bispectra are formulated in the same form as
the scalar one.  It is because the helicity dependence, which
brings nontrivial couplings between angular momenta in the reduced
bispectra, vanishes in the CMB 1-point functions induced from the
tensor and vector perturbations due to the absence of the contribution
of azimuthal angle from $\bf k$ to $\hat{n}$ in the transfer functions as
discussed in Appendix \ref{appen:flat_al}. Hence, unlike the all-sky
analysis, the sum of the reduced bispectrum is not needed in calculating
the tensor and vector bispectra in the flat-sky limit and
one can calculate the tensor and vector bispectra
with the same computational cost taken in the scalar case.
% correspondence to all sky
This corresponds to the
restoration of the orthogonality of $\theta_k$ between $Y_{\ell' m'}$
and ${}_{s}Y_{\ell m}$ ($s=\pm 1$ or $\pm 2$) for $\ell \gg 1$, namely,
$\mathcal{Y}^{(Z)}_{\ell \ell' m} (s) \rightarrow \delta_{\ell_1 \ell'_1} \delta_{\ell_2 \ell'_2} \delta_{\ell_3 \ell'_3}$ (there is no dependence on $m_1, m_2$ and $m_3$)
 due to
${}_s Y_{\ell m} \rightarrow Y_{\ell m}$ for $\ell \gg 1$.
In other words, it means that  because the
degeneracy factor of $m$ equal to $2 \ell + 1$ becomes so large in the large $\ell$ limit, the spin
eigenstate of $s = \pm 1$ or $\pm 2$ and that of $s = 0$ are almost indistinguishable.

%parity
In addition, interestingly, from Eq. (\ref{eq:flat_red_bi_TV}) and (\ref{eq:local_PNG}), we find that $B$-mode bispectra ($x = 1$) from tensor and
vector perturbations vanish if each $f_{Z,s_1 s_2}^{s_3}$ is identical
value. This situation corresponds to the parity conservation at
the nonlinear level. Therefore, if one detects the finite value of
the $B$-mode bispectrum in the squeezed limit, it may offer further evidence of the cosmological
parity violation.

%%%%%%%%%%%%%%%%%%%%%%%%%%%%%%%%%%%%%%%%%
\section{Summary and Discussion} \label{sec:sum}

In this paper, we derive the complete set of CMB temperature and
polarization bispectra generated from non-Gaussianity in the tensor and
vector-mode perturbations both in the all and flat-sky analyses.
For the primordial non-Gaussianity in the tensor and
vector sectors, we consider the more general type such as
Eq. (\ref{eq:general_PNG}), which contains the squeezed type given by Eqs. (\ref{eq:def_xi}) and (\ref{eq:def_xi_NL}) and the equilateral type. 

Note that the formulation presented can be
easily extended in a straightforward manner to the other cases, such as
a case in which the nonlinear tensor perturbation is excited by the
linear-order scalar-tensor couplings. 
As an example of this, we can consider the scalar-graviton interaction during inflation
shown by Ref.~\cite{Maldacena:2002vr}.
Through such interaction, the non-Gaussianity of the primordial fluctuations
can be generated as a scalar-scalar-tensor type, namely $\Braket{a^{(S)} a^{(S)} a^{(T)}}$.
Although, in the standard slow-roll inflation, such type of non-Gaussianity
is expected to be suppressed by the slow-roll parameter, it seems interesting that one investigates such type of non-Gaussianity through the future CMB
observations in the sense of the confirmation of the standard
inflation scenario, by using our formulation.
Furthermore, the 3-point cross
correlations between CMB intensity and polarizations, such as $\braket{a_I a_I
a_E}$, and higher-order correlations than the 3-point one can be easily formulated in the same manner~\cite{Munshi:2010bh, Okamoto/Hu:2002}.

In the formulation of all-sky bispectra, we find that those formulae take
complicated forms compared to the scalar one due to the
helicity dependence which is represented by the azimuthal angle dependence
between the wave vector of photon and the unit vector specifying the
line of sight direction in the photon propagation. However, in the
formulation of flat-sky bispectra, we find that the above difficulty is
solved for the absence of the above azimuthal dependence. 
In addition,
we also show that if the bispectra of $B$-mode polarization are a nonzero
value, it may become evidence of the cosmological parity
violation in the nonlinear sector.

\begin{acknowledgments}
This work is supported by Grant-in-Aid for JSPS Research under Grant No. 22-7477 (M. S.), and JSPS Grant-in-Aid for Scientific
Research under Grant Nos. 22340056 (S. Y.), 21740177 (K. I.), and 21840028 (K. T.).
This work is also supported in part by the Grant-in-Aid for Scientific
Research on Priority Areas No. 467 "Probing the Dark Energy through an
 Extremely Wide and Deep Survey with Subaru Telescope" and by the
 Grant-in-Aid for Nagoya University Global COE Program, "Quest for
 Fundamental Principles in the Universe: from Particles to the Solar
 System and the Cosmos," from the Ministry of Education, Culture,
 Sports, Science and Technology of Japan. 
\end{acknowledgments}

%=========== Appendix ===============================
\appendix
\section{Radiation transfer functions}\label{appen:transfer}
Here we show the radiation transfer functions of
temperature mode $\Delta_I$ and two polarization modes $\Delta_Q,
\Delta_U$.  Transfer functions induced by the scalar and tensor
modes in a particular basis in which the wave vector of photons ${\bf k}$
is parallel to $z$ axis $\hat{z}$ are formulated in Refs.~\cite{Kosowsky:1996,
Zaldarriaga/Seljak:1997, Weinberg:2008} by the line of sight integral
method. For the vector case, the method of calculation can be obtained
in Ref.~\cite{Brown:2008}.  Based on the Stokes parameters as defined in
 Ref.~\cite{Pitrou:2008ak}, these are expressed as follows:
\begin{eqnarray}
%---- scalar --------
\Delta_I^{(S)}(\tau_0, {\bf k}, \Omega_{k,n}) 
&=&  \Phi({\bf k}) \int_0^{\tau_0} d\tau
 S_I^{(S)}(k,\tau) e^{-i \mu_{k,n} x}~, \label{eq:trans_st} \\
\Delta_{Q}^{(S)} (\tau_0, {\bf k}, \Omega_{k,n}) 
&=& (1-\mu_{k,n}^2) \Phi({\bf k}) \int_0^{\tau_0} d\tau
 S_P^{(S)}(k,\tau) e^{-i \mu_{k,n} x} ~, \label{eq:trans_sq}\\
%--------- tensor -----
 \Delta^{(T)}_{I}(\tau_0, {\bf k},\Omega_{k,n})
&=& (1 - \mu_{k,n}^2) 
 \left( e^{2i\phi_{k,n}}\xi^{+2}({\bf k}) + e^{-2i\phi_{k,n}}\xi^{-2}({\bf
 k}) \right) \int_0^{\tau_0} d\tau
 S_I^{(T)}(k, \tau) e^{- i \mu_{k,n} x}~, \label{eq:trans_tt} \\
%---
(\Delta^{(T)}_Q \pm i\Delta^{(T)}_U)(\tau_0, {\bf k},\Omega_{k,n})
&=& [(1 \mp \mu_{k,n})^2 e^{2i\phi_{k,n}}\xi^{+2}({\bf k}) + (1 \pm
\mu_{k,n})^2 e^{-2i\phi_{k,n}}\xi^{-2}({\bf
 k})] \nonumber \\
&& \times \int_0^{\tau_0} d\tau
 S_P^{(T)}(k, \tau)  e^{- i \mu_{k,n} x}~, \label{eq:trans_tqu} \\
%--- vector ------------
\Delta^{(V)}_{I}(\tau_0, {\bf k}, \Omega_{k,n}) &=& -i \sqrt{1 - \mu_{k,n}^2} (\xi^{+1}({\bf
 k}) e^{i\phi_{k,n}} + \xi^{-1}({\bf k}) e^{-i\phi_{k,n}})
 \int_0^{\tau_0} d\tau S_I^{(V)}(k,\tau) e^{- i \mu_{k,n} x}~, \label{eq:trans_vt} \\
(\Delta^{(V)}_Q \pm i\Delta^{(V)}_U)(\tau_0, {\bf k},\Omega_{k,n})
&=& \sqrt{1-\mu_{k,n}^2} [\mp(1 \mp \mu_{k,n}) \xi^{+1}({\bf
 k}) e^{i\phi_{k,n}} \pm (1\pm \mu_{k,n})\xi^{-1}({\bf k})
 e^{-i\phi_{k,n}}] \nonumber \\
&& \times \int_0^{\tau_0} d\tau S_P^{(V)}(k,\tau) e^{ - i \mu_{k,n} x}~, \label{eq:trans_vqu}
\end{eqnarray}
where $\Omega_{k,n} (\equiv
(\theta_{k,n}, \phi_{k,n}))$ denotes the orientation of the line of
sight direction $\hat{n}$ in a particular basis in which ${\bf k} || \hat{z}$,  $\mu_{k,n} \equiv \cos \theta_{k,n}$, 
and $S_I^{(Z)}(k,\tau)$ and $S_P^{(Z)}(k,\tau)$ are the $Z$-type source functions of the temperature and polarization fluctuations \cite{Hu/White:1997,
Zaldarriaga/Seljak:1997, Landriau/Shellard:2003}

 In order to estimate the 1-point function $a_{\ell m}$, one must
 construct the transfer functions for the arbitrary ${\bf k}$.  
 In other words, we want to obtain the transfer functions expressed
 by the arbitrary ${\bf k}$ (whose direction is denoted by $\Omega_k$) and
 $\hat{n}$ (denoted by $\Omega_n$)
 instead of $\Omega_{k,n}$.  To achieve this 
 we introduce the rotational matrix
\begin{eqnarray}
S(\Omega_k) 
\equiv \left( 
  \begin{array}{ccc}
  \cos\theta_k \cos\phi_k & -\sin\phi_k  & \sin\theta_k \cos\phi_k \\
 \cos\theta_{k} \sin\phi_{k}  &  {\rm cos}\phi_{k} & \sin\theta_{k} \sin\phi_{k} \\
 -\sin\theta_k & 0 & {\rm cos}\theta_{k}
  \end{array}
 \right)~,
\end{eqnarray}
which expresses the basis rotation that transforms $\hat{z} \parallel {\bf
k}$ to the arbitrary $\hat{z}$.  Then the relation between $\Omega_k,
\Omega_n$ and $\Omega_{k,n}$ is written:
\begin{eqnarray}
 \left(
  \begin{array}{c}
  {\rm sin}\theta_{n} {\rm cos}\phi_{n} \\
  {\rm sin}\theta_{n} {\rm sin}\phi_{n} \\
  {\rm cos}\phi_{n} 
  \end{array}
 \right)
&=& S(\Omega_k)
 \left(
  \begin{array}{c}
  {\rm sin}\theta_{k,n} {\rm cos}\phi_{k,n} \\
  {\rm sin}\theta_{k,n} {\rm sin}\phi_{k,n} \\
  {\rm cos}\phi_{k,n} 
  \end{array}
 \right)
~. \label{eq:transform}
\end{eqnarray}
 In the temperature modes, only by changing $\Omega_{k,n}$ to $\Omega_k$
 and $\Omega_n$ with the relation (\ref{eq:transform}), the transfer
 functions for the arbitrary $\bf k$ can be obtained.
In the $E,B$ modes, in addition to this treatment, one must consider the
 mixing between $\Delta_Q$ and $\Delta_U$ under the transformation
 $S(\Omega_k)$ as described in Ref.~\cite{Zaldarriaga/Seljak:1997}. 

This effect is expressed as
\begin{eqnarray} 
(\Delta_Q' \pm i\Delta_U')(\tau_0, {\bf k}, \Omega_n) 
= e^{\mp 2i \psi} (\Delta_Q \pm i\Delta_U)(\tau_0, {\bf k},
\Omega_{k,n})~.
\end{eqnarray}
with the mixing angle $\psi$. The angle $\psi$ represents the rotation
angle between $\hat{\theta}_{k,n}$ and $\hat{\theta}_{n}$, where
$\hat{\theta}_{k,n}$ and $\hat{\theta}_{n}$ are the unit vectors orthogonal to $\hat{n}$ in
a particular basis in which ${\bf k}\parallel \hat{z}$ and a general basis, respectively.

In the
 flat-sky analysis, i.e., $\theta_n \rightarrow 0$, by using
 Eqs. (\ref{eq:trans_st}) - (\ref{eq:trans_vqu}) and (\ref{eq:transform})
 and by using the limit of $\psi$ as $\psi \rightarrow \phi_n -\phi_k + \pi$, the
 transfer functions for the arbitrary ${\bf k}$ are derived as
\begin{eqnarray}
%scalar------
\Delta_{I}^{(S)} (\tau_0, {\bf k}, \Omega_n) 
&\rightarrow& \Phi ({\bf k})  \int_0^{\tau_0} d\tau  S_I^{(S)}(k,\tau)
e^{- i {\bf k} \cdot \hat{n} D} ~, \label{eq:flat_trans_st}\\ 
(\Delta_Q^{(S)} \pm i\Delta_U^{(S)})(\tau_0, {\bf k}, \Omega_{n})
&\rightarrow& e^{\mp 2i (\phi_n -\phi_k)} \sin^2 \theta_k  \Phi({\bf k}) 
\int_0^{\tau_0} d\tau S_P^{(S)}(k,\tau) e^{- i {\bf k} \cdot \hat{n} D} ~,
\label{eq:flat_trans_squ} \\
%tensor-----
\Delta_{I}^{(T)} (\tau_0, {\bf k}, \Omega_n) 
&\rightarrow& (1-\mu^2_k)
\left( \xi^{+2} + \xi^{-2} \right )({\bf k})  \int_0^{\tau_0} d\tau
S_I^{(T)}(k,\tau) e^{-i {\bf k} \cdot \hat{n} D}~,
\label{eq:flat_trans_tt} \\
(\Delta_Q^{(T)} \pm i\Delta_U^{(T)})(\tau_0, {\bf k}, \Omega_n)
&\rightarrow& e^{\mp 2i (\phi_n - \phi_k)} 
\left[ (1 + \mu_k^2) (\xi^{+2}+\xi^{-2})({\bf k}) \mp 2 \mu_k
 (\xi^{+2}-\xi^{-2})({\bf k}) \right] \nonumber \\
&& \times \int_0^{\tau_0} d\tau S_P^{(T)}(k,\tau) e^{- i {\bf k} \cdot
 \hat{n} D} ~, \label{eq:flat_trans_tqu} \\
%vector--------
\Delta_{I}^{(V)} (\tau_0, {\bf k}, \Omega_n) 
&\rightarrow& i \sin \theta_k
\left( \xi^{+1} + \xi^{-1} \right )({\bf k}) \int_0^{\tau_0} d\tau
 S_I^{(V)}(k,\tau) e^{-i {\bf k} \cdot \hat{n} D}~,
 \label{eq:flat_trans_vt} \\
(\Delta_Q^{(V)} \pm i\Delta_U^{(V)})(\tau_0, {\bf k}, \Omega_n)
&\rightarrow& e^{\mp 2i (\phi_n - \phi_k)} 
\sin\theta_k \left[- \cos\theta_k (\xi^{+1}+\xi^{-1})({\bf  k}) \pm  (\xi^{+1}-\xi^{-1})({\bf k}) \right] \nonumber \\
&& \times \int_0^{\tau_0} d\tau S_P^{(V)}(k,\tau) e^{-i {\bf k} \cdot
 \hat{n} D}~. \label{eq:flat_trans_vqu}
\end{eqnarray} 
It is important to note that the $\phi_k$ dependence which are inherent
in the vector and tensor perturbations vanishes in the flat-sky
approximation, besides a trivial $\phi_k$ dependence due to a spin-2
nature of the Stokes $Q$ and $U$ parameters. One may explicitly see that
$\phi_{k,n}$ dependence vanishes in the transfer functions when taking
$\theta_n \rightarrow 0$ because the $S$ matrix rotates the basis with the
new $z$ axis always being on the $x-z$ plane in a particular basis in
which ${\bf k}\parallel \hat{z}$.  This approximation means that
for $\theta_n \ll 1$, it is valid to calculate the CMB fluctuation on
the basis of vector and tensor perturbations fixed as $\theta_n = 0$,
namely, $\phi_{k,n} = \pi$.
%--------
%%%%%%%%%%%%%%%%
\section{1-point function in the all-sky analysis} \label{appen:all_al}
% tensor temperature polarization
Here we formulate the all-mode 1-point functions
$a_{\ell m}$ in the all-sky analysis based on the derivation in
 Ref.~\cite{Weinberg:2008}. 
One-point functions of the $I,E,B$ modes are generated from $\Delta_I,\Delta_Q,\Delta_U$ as 
\begin{eqnarray}
a_{I, \ell m} &=& \int d\Omega_n \int \frac{d^3{\bf k}}{(2 \pi)^3}
 \Delta_{I}(\tau_0, {\bf k}, \Omega_n) Y_{\ell m}^*(\Omega_n)~,
 \label{eq:form_all_atlm} \\
%-----
a_{E, \ell m} &=& - \frac{1}{2} \int d\Omega_n \int \frac{d^3{\bf k}}{(2 \pi)^3} 
\left[ (\Delta_{Q} + i \Delta_U)(\tau_0, {\bf k}, \Omega_n)
 {}_{2}Y_{\ell m}^*(\Omega_n) + (\Delta_{Q} - i \Delta_U)(\tau_0, {\bf k},
 \Omega_n)  {}_{-2}Y_{\ell m}^*(\Omega_n) \right] \nonumber \\
&=& - \frac{1}{2} \left[ \frac{(\ell-2)!}{(\ell+2)!}\right]^{1/2} 
 \int d\Omega_n \int \frac{d^3{\bf k}}{(2 \pi)^3} 
\left[ \down^2 (\Delta_{Q} + i \Delta_U)
+ \up^2 (\Delta_{Q} - i \Delta_U) \right] (\tau_0, {\bf k}, \Omega_n) Y_{\ell m}^*(\Omega_n)~,
\label{eq:form_all_aelm} \\
%-------
a_{B, \ell m} &=& \frac{i}{2} \int d\Omega_n \int \frac{d^3{\bf k}}{(2 \pi)^3}
\left[ (\Delta_{Q} + i \Delta_U)(\tau_0, {\bf k}, \Omega_n)
 {}_{2}Y_{\ell m}^*(\Omega_n) - (\Delta_{Q} - i \Delta_U)(\tau_0, {\bf k},
 \Omega_n)  {}_{-2}Y_{\ell m}^*(\Omega_n) \right] \nonumber \\
&=& \frac{i}{2} \left[ \frac{(\ell-2)!}{(\ell+2)!}\right]^{1/2} 
 \int d\Omega_n \int \frac{d^3{\bf k}}{(2 \pi)^3} 
\left[ \down^2 (\Delta_{Q} + i \Delta_U)
- \up^2 (\Delta_{Q} - i \Delta_U) \right] (\tau_0, {\bf k}, \Omega_n)
Y_{\ell m}^*(\Omega_n)~. \label{eq:form_all_ablm}
\end{eqnarray}
Here we expand with the spin raising (lowering) operators $\up$
($\down$) as introduced in Refs.~\cite{Newman/Penrose:1966, Goldberg/etal:1967, Zaldarriaga/Seljak:1997} 
and $Y_{\ell m}(\Omega_n)$ for being easily understanding that $E,B$
modes are spin-0 fields. $\up$ and $\down$ act the spin-$s$ function ${}_s
f(\theta_n,\phi_n)$ as
\begin{eqnarray}
\up {}_s f(\theta_n,\phi_n) &=& -\sin^s \theta_n
\left[ \partial_{\theta_n} + i\csc \theta_n \partial_{\phi_n} \right]
\sin^{-s} \theta_n {}_s f(\theta_n,\phi_n)~, \\
\down {}_s f(\theta_n,\phi_n) &=& -\sin^{-s} \theta_n
\left[ \partial_{\theta_n} - i\csc \theta_n \partial_{\phi_n} \right]
\sin^{s} \theta_n {}_s f(\theta_n,\phi_n)~.
\end{eqnarray}

From here, we derive the 1-point function of tensor-temperature mode as
an example.  As mentioned in Sec. \ref{sec:all_bi}, This is calculated
by using Wigner $D$-matrix $D_{m m'}^{(\ell)}$, which is the unitary
irreducible matrix of rank $2\ell+1$ that forms a representation of the
rotational group. The property of this matrix and the relation with
spin-weighted spherical harmonics are explained in Refs.~\cite{Weinberg:2008,
Okamoto/Hu:2002, Goldberg/etal:1967}. By using
Eq. (\ref{eq:form_all_atlm}), the relation between the $Y_{\ell m}$ and
$D$ matrix,  and the relation corresponding to Eq. (\ref{eq:transform}) as
\begin{eqnarray}
Y_{\ell m}^*(\Omega_n)  &=& \sum_{m'} D_{m m'}^{(\ell)}
\left( S(\Omega_k) \right) Y_{\ell m'}^*(\Omega_{k,n})~, \\  
d \Omega_{n} &=& d \Omega_{k,n}~,
\end{eqnarray}
the 1-point function of tensor-temperature mode is written as
\begin{eqnarray}
a^{(T)}_{I,\ell m} 
= \int \frac{d^3{\bf k}}{(2 \pi)^3} \left[ \sum_{m'} D^{(\ell)}_{m m'}
\left( S(\Omega_k) \right) \int d \Omega_{k,n} Y^*_{\ell m'}(\Omega_{k,n})
\Delta^{(T)}_{I}(\tau_0, {\bf k},\Omega_{k,n}) \right]~. 
\end{eqnarray}
Next, with the mathematical relations as
\begin{eqnarray}
Y^*_{\ell m'}(\Omega_{k,n}) &=& \left[ \frac{2l+1}{4\pi} \frac{(\ell-m')!}{(\ell+m')!} \right]^{1/2}
 P_\ell^{m'} (\mu_{k,n}) e^{-i m' \phi_{k,n}} ~, \\
P_\ell^{-2}(\mu_{k,n}) &=& \frac{(\ell-2)!}{(\ell+2)!}P_\ell^2(\mu_{k,n})~,  \\
\int_{-1}^1 d\mu_{k,n} (1 - \mu_{k,n}^2) P_\ell^2(\mu_{k,n})  e^{-i\mu_{k,n}
 x} &=& -2
 (-i)^\ell \frac{(\ell+2)!}{(\ell-2)!} \frac{j_\ell(x)}{x^2} ~,
\end{eqnarray}
the integration for $\Omega_{k,n}$ can be performed to obtain
\begin{eqnarray}
 a^{(T)}_{I,\ell m} &=& -4 \pi (-i)^\ell 
\left[ \frac{(\ell+2)!}{(\ell-2)!} \right]^{1/2} 
\left[ \frac{2\ell + 1}{4 \pi} \right]^{1/2}
 \int \frac{d^3{\bf k}}{(2 \pi)^3} \left[ D^{(\ell)}_{m 2}
\left( S(\Omega_k) \right) \xi^{+2}({\bf k}) + D^{(\ell)}_{m,-2}
\left( S(\Omega_k) \right)
\xi^{-2}({\bf k}) \right] \nonumber \\
&& \times \int_0^{\tau_0} d\tau
 S_I^{(T)}(k, \tau) \frac{j_\ell(x)}{x^2}~. 
\end{eqnarray}
Because $D$ matrix is written by the spin-weighted spherical harmonics as
\begin{eqnarray}
D_{ms}^{(\ell)} \left( S(\Omega_{k}) \right) &=& 
\left[ \frac{4 \pi}{2\ell + 1} \right]^{1/2} (-1)^s
{}_{-s}Y_{\ell m}^*(\Omega_{k})~, \label{eq:d-matrix}
\end{eqnarray}
we obtain the final form, namely
\begin{eqnarray}
a^{(T)}_{I,\ell m} &=& -4 \pi (-i)^\ell 
\left[ \frac{(\ell+2)!}{(\ell-2)!} \right]^{1/2} \int \frac{d^3{\bf k}}{(2 \pi)^3} 
\left[{}_{-2}Y_{\ell m}^*(\Omega_k) \xi^{+2}({\bf k}) + {}_{2}Y_{\ell m}^*(\Omega_k) \xi^{-2}({\bf k}) \right] \int_0^{\tau_0} d\tau
 S_I^{(T)}(k, \tau) \frac{j_\ell(x)}{x^2}~. 
\end{eqnarray}

For the other modes, we can derive in the same manner with
Eqs.(\ref{eq:form_all_atlm}) - (\ref{eq:form_all_ablm}),
(\ref{eq:trans_st}), (\ref{eq:trans_sq}), (\ref{eq:trans_tqu}) -
(\ref{eq:trans_vqu}).

As a result, all-sky 1-point functions can be formulated:
\begin{eqnarray}
a^{(Z)}_{X, \ell m} &=& 
4\pi (-i)^{\ell} \int \frac{d^3{\bf k}}{(2 \pi)^3} \times
\begin{cases}
Y^*_{\ell m}(\Omega_k) \Phi({\bf k}) \mathcal{T}^{(Z)}_{X, \ell}(k) 
& ({\rm for \ } Z = S) \\
\sum_{s} {\rm sgn}(s)^{s+x} {}_{-s} Y^*_{\ell m}(\Omega_k) \xi^{s}({\bf k})
\mathcal{T}^{(Z)}_{X, \ell}(k)
& ({\rm for \ } Z = T,V) 
\end{cases}
\label{eq:all_alm_general} ~ ,
\end{eqnarray}
where $x = 0$ for $X=I,E$, $x = 1$ for $X=B$,
time-integrated transfer functions $\mathcal{T}^{(Z)}_{X, \ell}(k)$ are expressed as
\begin{eqnarray}
%----scalar
 \mathcal{T}_{I,\ell}^{(S)}(k) &=&  \int_0^{\tau_0} d\tau S_I^{(S)}(k,\tau)j_\ell(x)
  \label{eq:all_almst}~, \\
 \mathcal{T}_{E,\ell}^{(S)}(k) &=& \left[ \frac{(\ell-2)!}{(\ell+2)!} \right] ^{1/2} \int_0^{\tau_0} d\tau S_P^{(S)} \hat{\mathcal{E}}^{(S)}(x) j_\ell(x) \label{eq:all_almse}~, \\ 
%---- tensor 
\mathcal{T}_{I,\ell}^{(T)}(k) &=& - \left[ \frac{(\ell+2)!}{(\ell-2)!} \right]^{1/2} \int_0^{\tau_0} d\tau S_I^{(T)}(k, \tau) \frac{j_\ell(x)}{x^2} \label{eq:all_almtt}~, \\
 \mathcal{T}_{E,\ell}^{(T)}(k) &=& - \int_0^{\tau_0} d\tau S_P^{(T)}(k, \tau) \hat{\mathcal{E}}^{(T)}(x) \frac{j_\ell(x)}{x^2} \label{eq:all_almte}~, \\
 \mathcal{T}_{B,\ell}^{(T)}(k) &=& \int_0^{\tau_0} d\tau S_P^{(T)}(k, \tau) \hat{\mathcal{B}}^{(T)}(x) \frac{j_\ell(x)}{x^2} \label{eq:all_almtb}~, \\
%----- vector
\mathcal{T}_{I,\ell}^{(V)}(k) &=& - \left[ \frac{(\ell+1)!}{(\ell-1)!} \right]^{1/2}
\int_0^{\tau_0} d\tau S_I^{(V)}(k, \tau) \frac{j_\ell(x)}{x}
\label{eq:all_almvt}~, \\
 \mathcal{T}_{E,\ell}^{(V)}(k) &=&   
\left[ \frac{(\ell+1)!}{(\ell-1)!} \frac{(\ell-2)!}{(\ell+2)!} \right]^{1/2}
\int_0^{\tau_0} d\tau S_P^{(V)}(k, \tau) \hat{\mathcal{E}}^{(V)}(x)
\frac{j_\ell(x)}{x} \label{eq:all_almve}~, \\
 \mathcal{T}_{B,\ell}^{(V)}(k) &=& 
\left[ \frac{(\ell+1)!}{(\ell-1)!} \frac{(\ell-2)!}{(\ell+2)!} \right]^{1/2}
\int_0^{\tau_0} d\tau S_P^{(V)}(k, \tau) \hat{\mathcal{B}}^{(V)}(x)
\frac{j_\ell(x)}{x} \label{eq:all_almvb}~,
\end{eqnarray}  
and the operators $\mathcal{E}, \mathcal{B}$ are defined as 
\begin{eqnarray}
\begin{split}
\hat{\mathcal{E}}^{(S)}(x) &\equiv (1 + \partial_x^2)^2 x^2 ~, \\
\hat{\mathcal{E}}^{(T)}(x) &\equiv -12 + x^2 (1-\partial_x^2) - 8x
 \partial_x  ~, \\
\hat{\mathcal{B}}^{(T)}(x) &\equiv 8x + 2x^2 \partial_x  ~, \\
\hat{\mathcal{E}}^{(V)}(x) &\equiv 4x + (12+x^2)
 \partial_x + 8x \partial_x^2 + x^2 \partial_x^3  ~, \\
\hat{\mathcal{B}}^{(V)}(x) &\equiv x^2 + 4x\partial_x + x^2
 \partial_x^2~. \label{eq:x_operator}
\end{split}
\end{eqnarray} 
Note that in the all-sky analysis, due to the dependence of
transfer functions on $\phi_{k,n}$, 1-point functions depend on the
helicity state through the spin spherical harmonics.
%=============================================
\section{1-point function in the flat-sky analysis} \label{appen:flat_al}
In this section, we formulate the all-mode 1-point functions
$a_{\ell m}$ in the flat-sky analysis. 
In this limit, 1-point functions in the all-sky analysis
described as Eqs. (\ref{eq:form_all_atlm}) - (\ref{eq:form_all_ablm}) are
modified by using the plane wave as 
\begin{eqnarray}
a_{I, \ell m} &\rightarrow& \int d^2{\bf \Theta} \int \frac{d^3{\bf k}}{(2 \pi)^3} \Delta_{I}(\tau_0, {\bf k}, {\Omega_n}) e^{-i {\boldsymbol \ell} \cdot {\bf \Theta}} \equiv a_I({\boldsymbol \ell})~,  \label{eq:form_flat_atl} \\
a_{E, \ell m} &\rightarrow& \frac{1}{2} \int d^2{\bf \Theta} \int
 \frac{d^3{\bf k}}{(2 \pi)^3} \left[ (\Delta_{Q} + i \Delta_U) e^{-2i(\phi_\ell - \phi_n)} + (\Delta_{Q}
 - i \Delta_U) e^{2i(\phi_\ell - \phi_n)} \right](\tau_0, {\bf k}, {\Omega_n}) e^{-i {\boldsymbol \ell} \cdot {\bf \Theta}} \equiv a_E({\boldsymbol \ell}) \label{eq:form_flat_ael} ~, \\
a_{B, \ell m} &\rightarrow& \frac{i}{2} \int d^2{\bf \Theta} \int \frac{d^3{\bf k}}{(2 \pi)^3} 
\left[ - (\Delta_{Q} + i \Delta_U) e^{-2i(\phi_\ell - \phi_n)} +
 (\Delta_{Q} - i \Delta_U) e^{2i(\phi_\ell - \phi_n)} \right] (\tau_0, {\bf k}, {\Omega_n}) e^{-i {\boldsymbol \ell} \cdot {\bf \Theta}} \equiv a_B({\boldsymbol \ell})~, \label{eq:form_flat_abl}
\end{eqnarray} 
where $\bf \Theta$ is the 2D vector projecting $\hat{n}$ to the flat-sky
plane expressed as ${\bf \Theta} = (\Theta {\rm cos}\phi_n, \Theta
{\rm sin}\phi_n)$. For example, in order to obtain the 1-point function of the
tensor-temperature mode, we substitute Eq. (\ref{eq:flat_trans_tt}) into
Eq. (\ref{eq:form_flat_atl}) and calculate as follows:   
\begin{eqnarray}
a_I^{(T)}({\boldsymbol \ell}) &=& \int \frac{d^3{\bf k}}{(2 \pi)^3} (1-\mu^2_k)
\left( \xi^{+2} + \xi^{-2} \right ) ({\bf k})
\int_0^{\tau_0} d\tau \int d^2{\bf \Theta} e^{- i({\bf k}^{\parallel} D + {\boldsymbol \ell}) \cdot {\bf \Theta}} S_I^{(T)}(k,\tau) e^{-i k_z D} \nonumber \\
&=& \int \frac{d^3{\bf k}}{(2\pi)^3} \sin^2 \theta_k \left( \xi^{+2} + \xi^{-2} \right ) ({\bf k})
\int_0^{\tau_0} d\tau (2\pi)^2 \delta^{(2)}({\bf k}^{\parallel}D + {\boldsymbol \ell}) S_I^{(T)}(k,\tau) e^{-i k_z D} \nonumber \\
&=& \int_0^{\tau_0} d\tau \int_{-\infty}^{\infty} \frac{dk_z}{2\pi}
 (\xi^{+2} + \xi^{-2}) ({\bf k}^{\parallel} = -{\boldsymbol \ell}/D, k_z)
 \frac{\ell^2}{(k_z D)^2 + \ell^2} S_I^{(T)}(k = \sqrt{k_z^2 + (\ell/D)^2},\tau)
 \frac{1}{D^2}e^{-i k_z D} \label{eq:flat_altt}~,
\end{eqnarray} 
where $D=\tau_0-\tau$ is the conformal distance and we have decomposed
$\bf{k}$ into two-dimensional vector parallel to the flat sky and that
orthogonal to it, ${\bf k}=({\bf k}^{\parallel},k_z)$.
In order to obtain the last equation, we use following relations which are
satisfied under ${\bf k}^{\parallel} = -{\boldsymbol \ell}/D$ as
\begin{eqnarray}
\begin{split}
k &= \sqrt{k_z^2 + 
\left(\frac{\ell}{D} \right)^2}  ~, \\
\sin \theta_k &= \frac{\ell}{kD} = \frac{\ell}{\sqrt{(k_zD)^2 + \ell^2}}  ~, \\
\cos \theta_k &= {\rm sgn}(k_z)
\sqrt{1 - 
\left( \frac{\ell}{kD} \right)^2 }  ~, \\
\phi_k &= \phi_\ell + \pi ~. \label{eq:delta2_condi}
\end{split}
\end{eqnarray}
One-point functions of the other modes are calculated in the same manner
by using Eqs. (\ref{eq:flat_trans_st}), (\ref{eq:flat_trans_squ}),
(\ref{eq:flat_trans_tqu}) - (\ref{eq:flat_trans_vqu}) and
(\ref{eq:form_flat_atl}) - (\ref{eq:form_flat_abl}) as
\begin{eqnarray}
a_I^{(S)}({\boldsymbol \ell}) &=& \int_0^{\tau_0} d\tau \int_{-\infty}^{\infty}
 \frac{dk_z}{2 \pi} \Phi ({\bf k}^{\parallel} = -{\boldsymbol \ell}/D, k_z)
 S_I^{(S)}(k = \sqrt{k_z^2 + (\ell/D)^2},\tau) \frac{1}{D^2}e^{- i k_z D} ~,
 \label{eq:flat_alst} \\
a_E^{(S)}({\boldsymbol \ell}) &=& \int_0^{\tau_0} d\tau
 \int_{-\infty}^{\infty} \frac{dk_z}{2 \pi} \Phi ({\bf k}^{\parallel} =
 - {\boldsymbol \ell}/D, k_z) 
\frac{\ell^2}{(k_z D)^2 + \ell^2}
S_P^{(S)}(k = \sqrt{k_z^2 + (\ell/D)^2},\tau)
 \frac{1}{D^2}e^{- i k_z D}  ~, \label{eq:flat_alse} \\
a_E^{(T)}({\boldsymbol \ell}) &=& \int_0^{\tau_0} d\tau \int_{-\infty}^{\infty} \frac{dk_z}{2\pi}
 (\xi^{+2} + \xi^{-2}) ({\bf k}^{\parallel} = - {\boldsymbol \ell}/D, k_z)
 \nonumber \\ 
&& \times \left( 2 - \frac{\ell^2}{(k_z D)^2 + \ell^2} \right) S_P^{(T)}(k =
 \sqrt{k_z^2 + (\ell/D)^2},\tau) \frac{1}{D^2} e^{-i k_z D}  ~, \label{eq:flat_alte} \\
a_B^{(T)}({\boldsymbol \ell}) &=& i \int_0^{\tau_0} d\tau \int_{-\infty}^{\infty}
 dk_z (\xi^{+2} - \xi^{-2}) ({\bf k}^{\parallel} = - {\boldsymbol \ell}/D, k_z) \nonumber \\
&& \times 2 ~{\rm sgn}(k_z) \sqrt{ 1 - \frac{\ell^2}{(k_z D)^2 + \ell^2} }
 S_P^{(T)}(k = \sqrt{k_z^2 + (\ell/D)^2},\tau) \frac{1}{D^2}e^{-i k_z D} ~,
 \label{eq:flat_altb} \\
a_I^{(V)}({\boldsymbol \ell}) &=& i \int_0^{\tau_0} d\tau
 \int_{-\infty}^{\infty} \frac{dk_z}{2 \pi} (\xi^{+1} + \xi^{-1})
 ({\bf k}^{\parallel} = - {\boldsymbol \ell}/D, k_z) \nonumber \\
&& \times \frac{\ell}{\sqrt{(k_z D)^2 + \ell^2}} S_I^{(V)}(k = \sqrt{k_z^2 +
(\ell/D)^2},\tau) \frac{1}{D^2}e^{-i k_z D}  ~, \label{eq:flat_alvt} \\
a_E^{(V)}({\boldsymbol \ell}) &=& -\int_0^{\tau_0} d\tau \int_{-\infty}^{\infty}
 \frac{dk_z}{2\pi} 
(\xi^{+1} + \xi^{-1}) ({\bf k}^{\parallel} = -{\boldsymbol \ell}/D, k_z) \nonumber \\
&& \times {\rm sgn}(k_z) \frac{\ell}{\sqrt{(k_z D)^2 + \ell^2}} \sqrt{1 -
 \frac{\ell^2}{(k_z D)^2 + \ell^2}} S_P^{(V)}(k = \sqrt{k_z^2 + (\ell/D)^2},\tau)
 \frac{1}{D^2}e^{-i k_z D}  ~, \label{eq:flat_alve} \\
a_B^{(V)}({\boldsymbol \ell}) &=& -i \int_0^{\tau_0} d\tau \int_{-\infty}^{\infty} \frac{dk_z}{2\pi} 
(\xi^{+1} - \xi^{-1}) ({\bf k}^{\parallel} = -{\boldsymbol \ell}/D, k_z) \nonumber \\
&& \times \frac{\ell}{\sqrt{(k_z D)^2 + \ell^2}} S_P^{(V)}(k = \sqrt{k_z^2 + (\ell/D)^2},\tau) \frac{1}{D^2}e^{-i k_z D}~. \label{eq:flat_alvb}
\end{eqnarray} 
Note that the helicity dependence vanishes in flat-sky 1-point
functions unlike in the all-sky ones as shown in Appendix
\ref{appen:all_al}. It is due to the absence of $\phi_{k,n}$ dependence in the flat-sky transfer functions as explained
in Appendix \ref{appen:transfer}.
\bibliography{paper}
%\nocite{*}
\end{document}